\documentclass[twocolumn]{aastex6}

\pdfoutput=1

\newcommand{\floor}[1]{\lfloor #1 \rfloor}
\newcommand{\ceil}[1]{\lceil #1 \rceil}

\newcommand{\wwwcoolworlds}{\href{https://github.com/CoolWorlds/exoinformatics}{this URL}}
\newcommand{\python}{{\tt Python}}

\usepackage{amsmath}
\usepackage{amsfonts}
\usepackage{amssymb}
\usepackage{wasysym}
\usepackage{comment}

\begin{document}


\title{Do Planets Remember How They Formed?}


\author{
David Kipping\altaffilmark{1}
}

\affil{dkipping@astro.columbia.edu}

\altaffiltext{1}{Department of Astronomy, Columbia University, 550 W 120th St., New York, NY 10027}

\begin{abstract}
	
One of the most directly observable features of a transiting multi-planet
system is their size-ordering when ranked in orbital separation.
\textit{Kepler} has revealed a rich diversity of outcomes, from perfectly
ordered systems, like Kepler-80, to ostensibly disordered systems, like
Kepler-20. Under the hypothesis that systems are born via preferred formation
pathways, one might reasonably expect non-random size-orderings reflecting
these processes. However, subsequent dynamical evolution, often chaotic and
turbulent in nature, may erode this information and so here we ask - do systems
remember how they formed? To address this, we devise a model to define the
entropy of a planetary system's size-ordering, by first comparing differences
between neighboring planets and then extending to accommodate differences
across the chain. We derive closed-form solutions for many of the microstate
occupancies and provide public code with look-up tables to compute entropy for
up to ten-planet systems. All three proposed entropy definitions exhibit the
expected property that their credible interval increases with respect to a
proxy for time. We find that the observed \textit{Kepler} multis display a
highly significant deficit in entropy compared to a randomly generated
population. Incorporating a filter for systems deemed likely to be dynamically
packed, we show that this result is robust against the possibility of missing
planets too. Put together, our work establishes that \textit{Kepler} systems do
indeed remember something of their younger years and highlights the value of
information theory for exoplanetary science.

\end{abstract}

\keywords{astroinformatics --- combinatorics --- planetary systems}



\section{INTRODUCTION}
\label{sec:introduction}

Extrasolar planetary systems reveal a rich diversity of architectures, most of
which do not directly resemble our own (e.g. see
\href{http://exoplanetarchive.ipac.caltech.edu}{NEA}, \citealt{NEA}, and
\href{http://www.exoplanets.org}{exoplanets.org}, \citealt{exoplanetsorg}). The 
ensemble properties of exoplanetary systems has been frequently exploited as
a window into the mechanisms guiding their formation and evolution, for example
by analyzing planet-metallicity trends \citep{1997MNRAS.285..403G,
2005ApJ...622.1102F,2012Natur.486..375B,2013ApJ...767L..24D}, 
mutual inclinations \citep{2012AJ....143...94T,2012ApJ...761...92F,
2014ApJ...790..146F,2016ApJ...816...66B}, orbital eccentricities
\citep{2008ApJ...685..553S,2011MNRAS.418.1822W,2013MNRAS.434L..51K,
2015ApJ...808..126V,2016ApJ...820...93S} and host star correlations
\citep{2010PASP..122..905J,2014ApJ...788..148S,2015ApJ...809....8B}.

As the majority of exoplanets discovered have come from the \textit{Kepler
Mission} via the transit method \citep{keplermission}, these systems
are particularly useful for such studies since they provide a homogeneous
sample for analysis with well-known biases \citep{sandford2016}. Using
transits, the two most robust observables are the orbital period of the planet
(from the timing between consecutive transits) and the planet-to-star radius
ratio (from the depth of the transits). The relative sizes of planets, ordered
from shortest-period to longest-period, is therefore also one of the most
robust observational signatures of system architectures available to us.
An example exploitation of this information comes from \citet{ciardi2013}, who
demonstrated that neighboring planet pairs tend to have the larger planet
on the outside if one of them is Neptune-sized or larger.

Any study seeking to learn something about formation pathways using the
size-ordering must operate under the fundamental assumption that the
observed size-ordering contains some information about the formation condition.
This statement cannot be simply assumed to be true, since the architectures
we observe are the product not just specific formation pathways but also
subsequent evolution.

Planetary systems are not static but continuously evolve, both with their
disks \citep{1986ApJ...309..846L} and planetesimals
\citep{2005Natur.435..459T} at early times, but also subsequently over many Gyr
through secular and chaotic dynamical interactions \citep{2008ApJ...683.1207B,
2012ApJ...755L..21D}. Over a long enough time then, the latter effect will
ultimately erode any memory of the initial formation conditions. Accordingly,
there is a need to test whether this highly robust observable - the
\textit{Kepler} system size-orderings - actually retains any information (or 
equivalently any ``memory'') at all of its initial formation.

The question of information content naturally lends itself to the field of
information theory. If we were able to define a metric to quantify the
Shannon entropy \citep{shannon,shannon1949} of planetary size-orderings, one
should expect that for an ensemble of systems, the average entropy would
increase over time as they accumulate dynamical interactions. This is because
Shannon entropy increases with the number of ways of organizing a system into
unique microstates, and systems with freedom to move between microstates (in
our case via dynamics) will tend to evolve towards the most frequently occupied
states. The key to defining entropy here ultimately boils down to how one
defines such microstates.

We might immediately consider the Kolmogorov-Sinai (KS) entropy 
\citep{kolmogorov,sinai} as a possible solution, which directly relates to the
Lyapunov exponent for dynamical systems \citep{1983rsm..book.....L}.
However, the calculation of Lyapunov exponents requires knowledge of the
full system properties, including eccentricity and mutual inclination
which are not direct observables from a transit \citep{seager2003}.

An alternative entropy term for planetary architectures is given in
\citet{tremaine2015} (using the partition function in Equation~11). The
Tremaine-entropy describes the distribution of the planetary orbits in
phase space and thus is primarily controlled by the relative spacing or
packing of the system. A pure size-order based entropy term, which we
seek in this work, is both distinct and complementary to this as it
concerns itself with the distribution of radii across their rank-ordering.
Whilst the Tremaine-entropy is a continuous function primarily describing
dynamical packing, a size-ordered entropy is a discrete function relating
primarily to the formation process itself (sizes expected to be governed
by disk densities, heating, turbulence and condensation locations) modulated
by subsequent dynamical evolution.

In this paper, we explore several candidate definitions for a size-ordering
entropy in Sections~\ref{sec:entropyT}, \ref{sec:entropyI}, \&
\ref{sec:entropyC}. After establishing that all three are measures of disorder
and increase with respect to a proxy for time, we apply them to \textit{Kepler}
systems in Section~\ref{sec:kepler}.
\\
\\

\section{A SIMPLE ENTROPY USING TALLY-SCORES}
\label{sec:entropyT}

\subsection{Concept}
\label{sub:concept}

We begin by defining what we mean by ``entropy'' of a planetary system.
This work focusses on a Shannon entropy like definition in relation to the
size ordering of planetary system architectures, but one could equally
consider other axes such as semi-major axes or inclinations, for example.

\begin{figure*}
\begin{center}
\includegraphics[width=17.0cm,angle=0,clip=true]{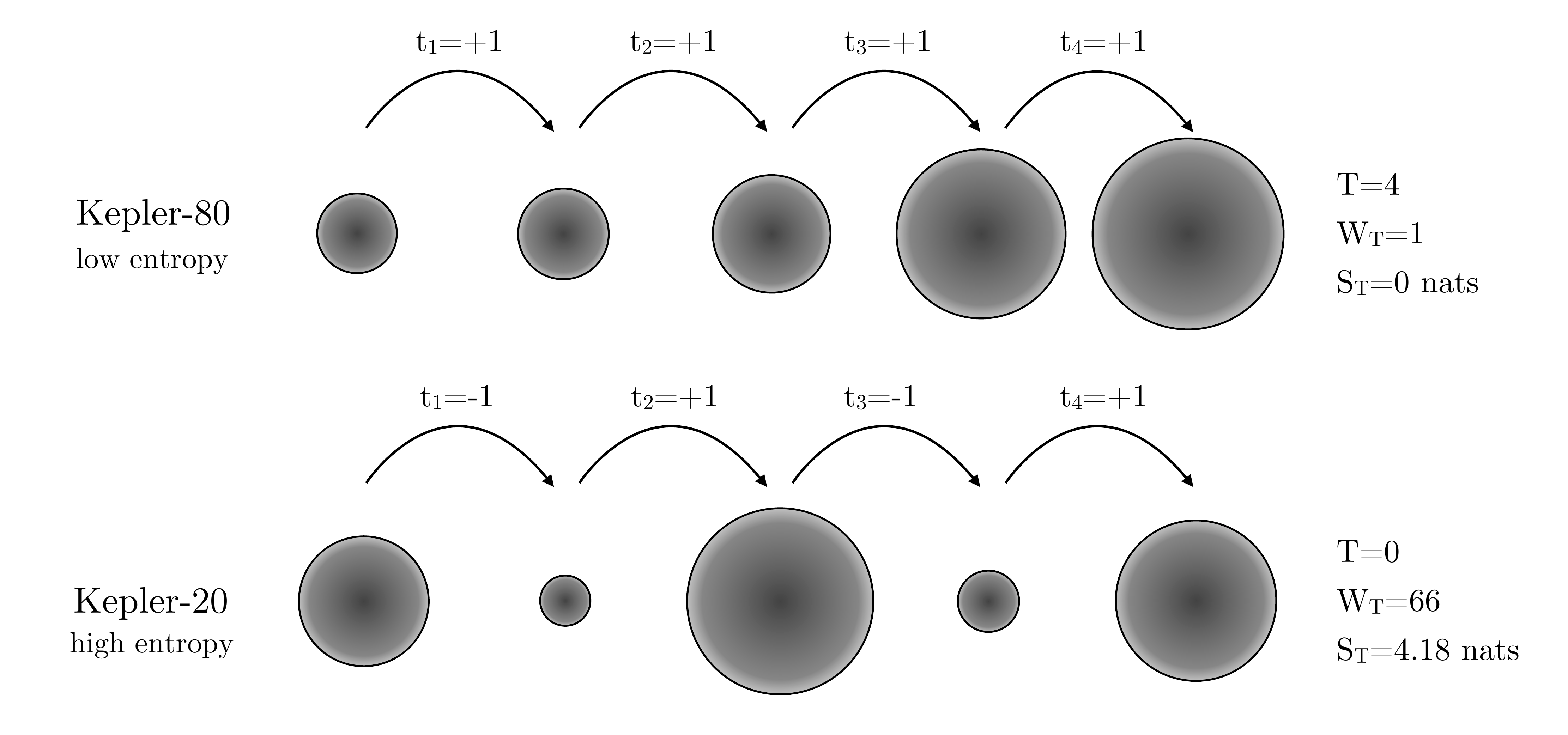}
\caption{
Example of the different planetary entropies for a five planet system.
The circles depict the size ordering of planetary radii, in order of
orbital separation from the star. The Kepler-80 planets appear sorted
in terms of their radii, giving a tally of $T=4$, which can only be 
achieved in this single configuration, thereby yielding a low entropy.
Vice versa, the Kepler-20 system is more disordered and high entropy.
Size orderings are based on the maximum likelihood planet-to-star relative
radii reported on the NASA Exoplanet Archive \citep{NEA}.
}
\label{fig:example}
\end{center}
\end{figure*}

The broad concept and thesis of this work is captured by the illustration
shown in Figure~\ref{fig:example}. Specifically, we were motivated to
devise a way of formally defining entropy for a system like Kepler-20
\citep{kepler20}, which displays an ostensibly randomized size-ordering of
the known transiting planets. The disorder of such a configuration is
suitable for describing using Shannon entropy. In contrast, the Solar System
appears to have a quasi-sequential size ordering up to Jupiter, reversing
from that point down to Neptune. In the absence of the hypothesized inner
disk truncation by Jupiter during the Grand Tack scenario, Mars may have
grown to a Super-Earth leading to a great degree of size-ordering
\citep{walsh2011}. Ultimately, we seek here a quantitative metric to describe
these differences, rather than simply eye-balling disorder.

\subsection{The Number of Unique Microstates}

As an initial definition for size-ordering entropy, we consider the changes
between neighboring planets only. We may compute a size-ordering tally,
$T$, for each system by going through each pair and assigning $t_i=+1$
if the outer planet is larger than the inner, and $t_i=-1$ otherwise. This 
simple algorithm is equivalent to the expression

\begin{align}
t_i &= 2\mathbb{H}[R_{i+1}-R_i] - 1,
\end{align}

which yields a total tally score of

\begin{align}
T &\equiv \sum_{i=1}^{N-1} t_i,
\end{align}

where $R_i$ is the radius of the $i^{\mathrm{th}}$ planet and $N$ denotes
the total number of planets in the system. One may show that the resulting
entropy term we later compute using this definition is equivalent to different
scoring schemes, such as replacing $t_i=-1$ with $t_i=0$.

Each score can be considered to be a micro-state, and in total there will be
$N!$ possible microstates/scores and $\Omega_T$ unique micro-states/scores,
where

\begin{align}
\Omega_T &= N.
\end{align}

To distinguish between unique microstates, we label each with the index $M$, such
that  $M=1,2,...,N-1,N$.

\subsection{Determining Microstate Index}
\label{sub:whereami}

Whilst we have defined the number of unique microstates for $N$ planets,
$\Omega_T$, the actual score ($T$) of each microstate is yet to be defined.
One may show that the tally score of each unique microstates,
$T$, will follow an arithmetic series, given by

\begin{align}
T &= -(N-1) + 2(M-1).
\end{align}

Using the above, one can re-arrange to solve for $M$, the unique microstate
index, for a given solution for $T$ and a choice of $N$, to yield

\begin{align}
M &= \frac{N+T+1}{2}.
\label{eqn:orientT}
\end{align}

Equation~(\ref{eqn:orientT}) essentially allows one to orient ourselves and
convert an observed tally score into a location, in terms of unique microstate
index.

\subsection{Occupancies}

Let us define the occupancy of each unique microstate as $W_T$.
The minimum occupancy of a unique microstate (or the minimum frequency of
a unique score) is always 1. For $N>1$, there are always exactly two
unique microstates with an occupancy of $W_T=1$, which occur for
$T_{\mathrm{max}}=(N-1)$ and $T_{\mathrm{min}}=-(N-1)$.

Consider ranking all of the unique microstates from smallest to largest
$T$ and then labeling these ranks $M=1,2,...,N-1,N$ consecutively. The
occupancy, $W_T$, of the $M^{\mathrm{th}}$ unique microstates is described
by

\begin{align}
W_T &= A_{N,M-1},
\label{eqn:WTtemp}
\end{align}

where $A_{p,q}$ is the Eulerian number generating function given by

\begin{align}
A_{p,q} &\equiv \sum_{j=0}^q (-1)^j \binom{p+1}{j} (q-j+1)^p.
\end{align}

If $N$ is odd, then there will be just one unique microstate maximally occupied
with $T=0$. Else, if $N$ is even, there will a pair of unique
microstates/scores with maximal occupancy, given by $T=\{-1,+1\}$.

Combining our earlier result for $M$ as a function of $T$
(Equation~\ref{eqn:orientT}), with Equation~(\ref{eqn:WTtemp}) above,
allows us to directly evaluate $W_T$ from the score, $T$, using

\begin{align}
W_T &= A_{N,\tfrac{N+T-1}{2}}.
\end{align}

\begin{figure*}
\begin{center}
\includegraphics[width=17.0cm,angle=0,clip=true]{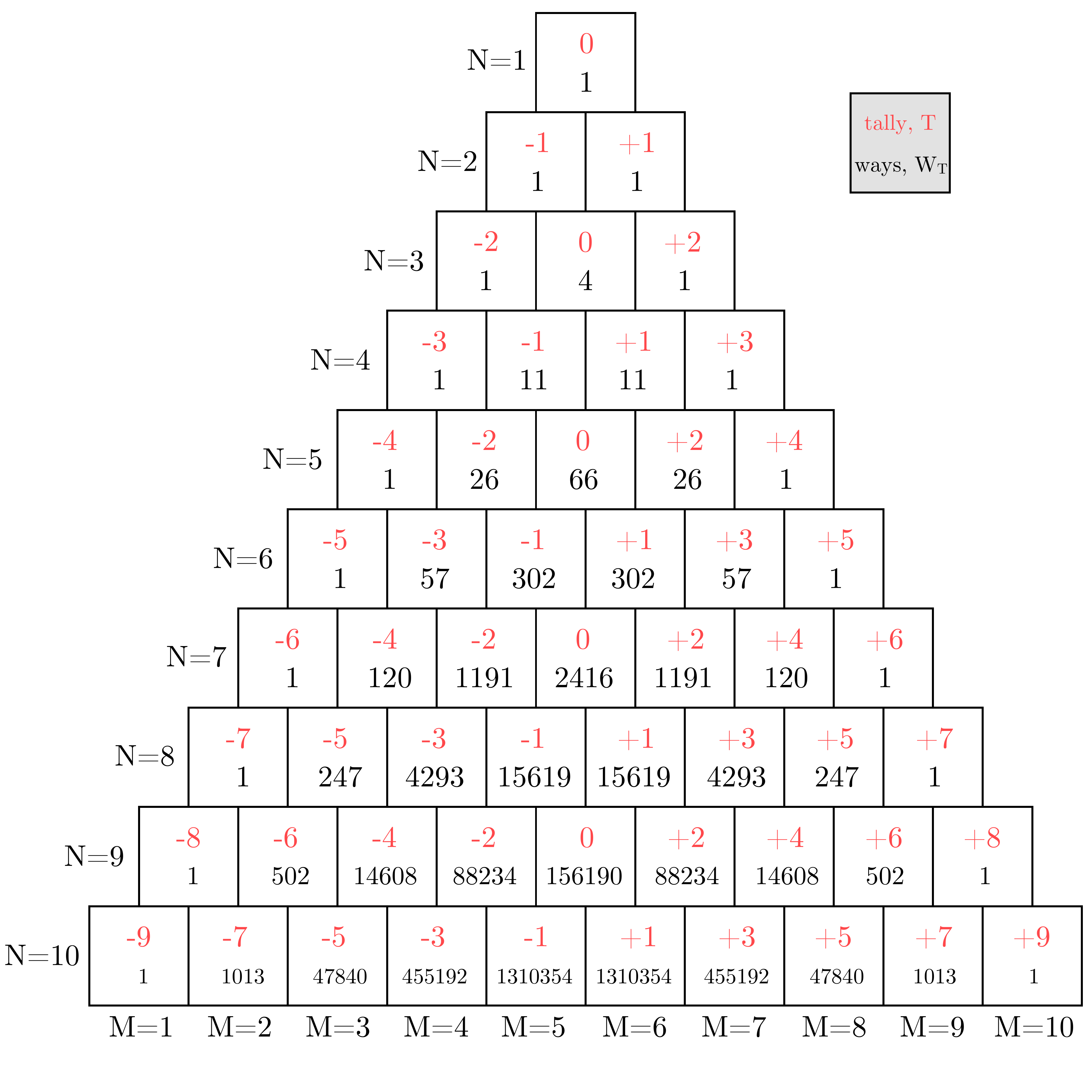}
\caption{
Illustration of the number of unique microstates for each choice of $N$.
Each unique microstate is represented by a black box, with the red
numbers giving the score, $T$, associated with that unique microstate and the
black numbers giving the number of microstates which can produce this
score, $W_T$.
}
\label{fig:microstates}
\end{center}
\end{figure*}

The scores and occupancies of each unique microstate, $T$ and $W_T$,
are illustrated in Figure~\ref{fig:microstates} up to $N=10$ (although
one can extend our approach to arbitrarily large $N$).

The occupancy of each unique microstate can be considered to be the number of
equivalent ways of obtaining the same score. In this way, we may define the
entropy of each unique microstate as 

\begin{align}
S_T = \log W_T,
\label{eqn:sizeentropy}
\end{align}

where we use log base $e$ in what follows to give an entropy in units of nats.

The minimum entropy obtainable thus occurs for the maximum/minimum score,
which has an occupancy of just one and thus $S_{T,\mathrm{min}}=\log 1 = 0$.
The maximum entropy obtainable occurs for the most occupied microstate, which
has an occupancy of $W_{T,\mathrm{max}}=A_{N,\floor{N/2}}$ and thus
$S_{T,\mathrm{max}} = \log A_{N,\floor{N/2}}$. There are only $\ceil{N/2}$
unique possible entropies, which is less than or equal to the number of unique
microstates, $N$ (since the $N^{\mathrm{th}}$ entropy vector is symmetric about
the median element).

\subsection{Evolving Entropy}
\label{sub:evolve}

A basic expectation of any definition of entropy is that it should increase
over time. To investigate this, we generated a initially pristine system of
$S_T=0$ such that the size orderings can be described by a vector containing an
integer sequence. Given that our entropy is framed in terms of size-ordering,
a suitable proxy for time would to allow exchanges between planets, for which
local neighboring pairs would be the simplest method. To accomplish this, we
chose a random element of the radii vector, followed by a random neighbor
either preceding or proceeding the element. For end-members, there is only one
choice for this element choice. The two elements are then exchanged, giving rise
to a higher entropy system. We repeated this $10^6$ times with an exchange
probability of $0.1$, leading to a final state which has been extremely well-mixed.

Although our algorithm only allows neighbors to swap, real physical systems
may be able to exchange non-neighboring planets too. However, such exchanges
can always be described by multiple neighboring swaps. We stress that the goal
of this exercise is not to find a causal relation between iteration number and
time, merely to qualitatively simulate the passage of time by allowing
exchanges to occur at some finite rate.

\begin{figure}
\begin{center}
\includegraphics[width=8.4cm,angle=0,clip=true]{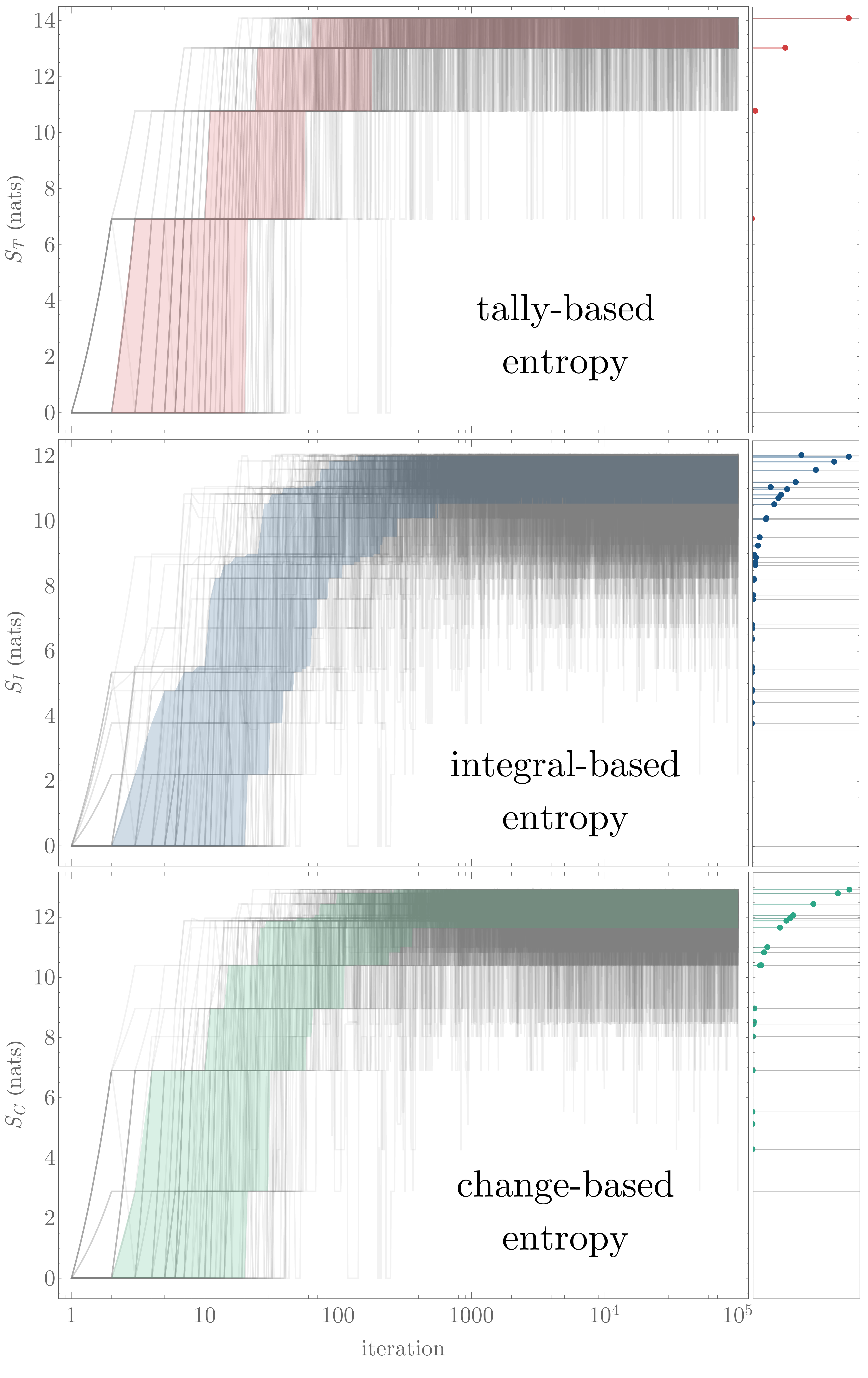}
\caption{
Evolution of the entropy of an initially pristine $N=10$ system where each
iteration allows for a 10\% probability swap of a random, neighboring
planet-pair. The 100 faint gray lines show individual simulations,
whereas the shaded region shows the 68.3\% central quantile. The
histogram on the right-hand side is that after $10^6$ iterations.
}
\label{fig:history}
\end{center}
\end{figure}

As shown in Figure~\ref{fig:history}, the average entropy of the system indeed
evolves ever-upwards, as expected for a entropy-like term. This establishes
that specific examples of low-entropy systems, such as Kepler-80 \citep{kepler80},
are highly unlikely to be the product of a purely random process. If their
size-ordering configuration are not random, then this implies that they retain
some information about a specific mechanism leading to their origin. In other
words, such low-entropy systems are information-rich.

\section{ENTROPY USING THE INTEGRAL PATH}
\label{sec:entropyI}

\subsection{The Solar System Counter Example}

\begin{figure*}
\begin{center}
\includegraphics[width=14.0cm,angle=0,clip=true]{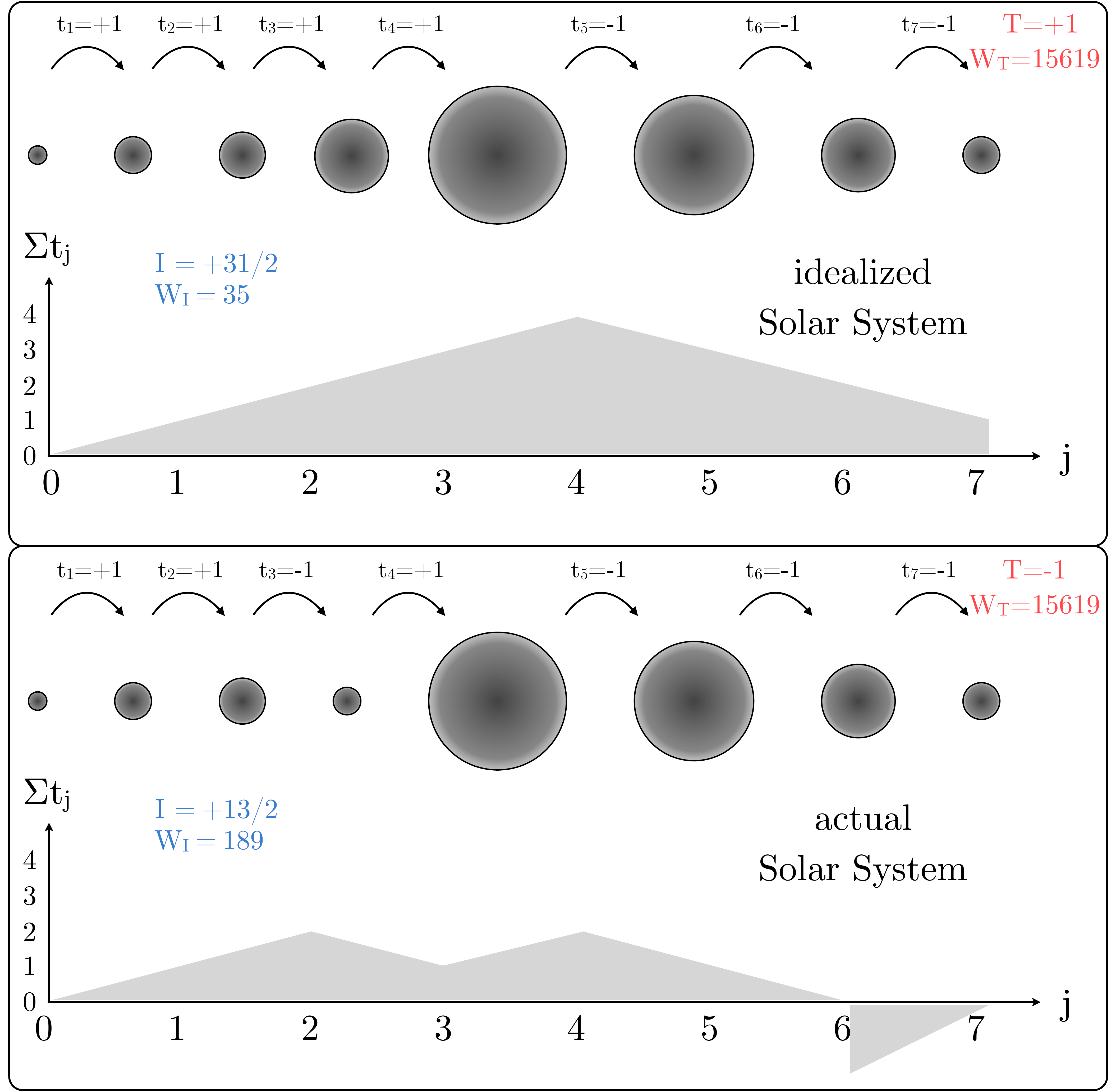}
\caption{
Illustration of the size ordering of the Solar System (not to scale)
for the actual configuration and an ``idealized'' configuration replacing
Mars with a Super-Earth. A tally-based entropy fails to detect the
low entropy nature of even the idealized state, motivating us to define
an improved definition using the integral-paths and change points, as
depicted.
}
\label{fig:Solar}
\end{center}
\end{figure*}

Although our definition of entropy appears to be characterizing disorder, the
Solar System provides a counter example as to why our current definition is
somewhat limited. For the $N=8$ planet Solar System, we find a score of
$T=-1$, which has $W_T=15619$ ways of achieving (see Figure~\ref{fig:Solar}).
Indeed, even replacing Mars with a Super-Earth, to create an apparently
highly-ordered size ordering leads to the same entropy. Thus, the Solar System
appears to be in the most highly disordered state possible. The clear size
ordering trend present (Figure~\ref{fig:Solar}) elucidates that our definition
is somehow inadequate and we consider why here.

First, consider that the Solar System planets increase in size from the inner
most planet, Mercury, up to Jupiter (with the exception of Mars), leading to a
high tally by the time we reach Jupiter. After this point, the planets
regularly decrease in size down to Neptune. Consequently, the high positive score
attained up to Jupiter is cancelled out by the high negative score of the outer
Solar System. Accordingly, this configuration scores the same as randomly
mixing the planets up.

This point illustrates what is wrong with our current definition for the score
- it does not have a memory. A score which flips randomly from one planet to
the next should be more disordered than a regular increase followed by a
regular decrease. We thus need to adapt our entropy definition to include a
term accounting for the autocorrelation or memory of the previous scores.

We devised two modifications to our simple tally system, one based on the
integral path and the other based on change points. We discuss the latter later
in Section~\ref{sec:entropyC}, and consider the integral path modification in
what follows.
\\

\subsection{Number of Unique Sub-Microstates}

Instead of considering each microstate as solely defined by the tally, $T$, we
consider that it is defined by a vector containing two numbers, $T$ and another
term chosen to incorporate a memory-like property. This much like how a
thermodynamic microstate can be defined by two degrees of freedom, rather than
just one for example. One way to think about this is that we have split the
original $M^{\mathrm{th}}$ microstate into $\omega_{I,M}$ ``sub-microstates'',
denoted by the labels $k=1,2,...,\omega_{I,M}-1,\omega_{I,M}$. The term
``sub-microstates'' is formally incorrect, but it useful since it allows us to
refer to these microstates relative to the original microstates derived using
the tally-based system. For this reason, we will use the term in what follows
but stress it is only for linguistic convenience.

A possible choice for a second degree of freedom is the integral of $t_i$ over
$i$. For example, $N$ consecutive increases in $t_i$ can be thought of a line
enclosing a triangular area below the curve (a linear interpolation). Let us
define this integral as $I$, which we use as the second term in a vector, such
that each unique sub-microstate is now defined by $\{T,I\}$.

We find that this definition is effective at splitting up the originally
defined microstates into a finer grid of sub-microstates and also brings the
Solar System down into a much lower entropy state, as desired (as shown later).

We find that the number of unique microstates grows from $\Omega_T=N$ to
$\Omega_I=\mathcal{C}_{N-1}$ sub-microstates, where $\mathcal{C}_k$ is a Cake
number, the maximum number of regions which a three-dimensional cube can be
partitioned by exactly $N$ planes, defined by

\begin{align}
\mathcal{C}_k &= \sum_{i=0}^k \binom{k}{i} = \frac{k^3 + 5 k + 6}{6}.
\end{align}

\begin{figure}
\begin{center}
\includegraphics[width=8.4cm,angle=0,clip=true]{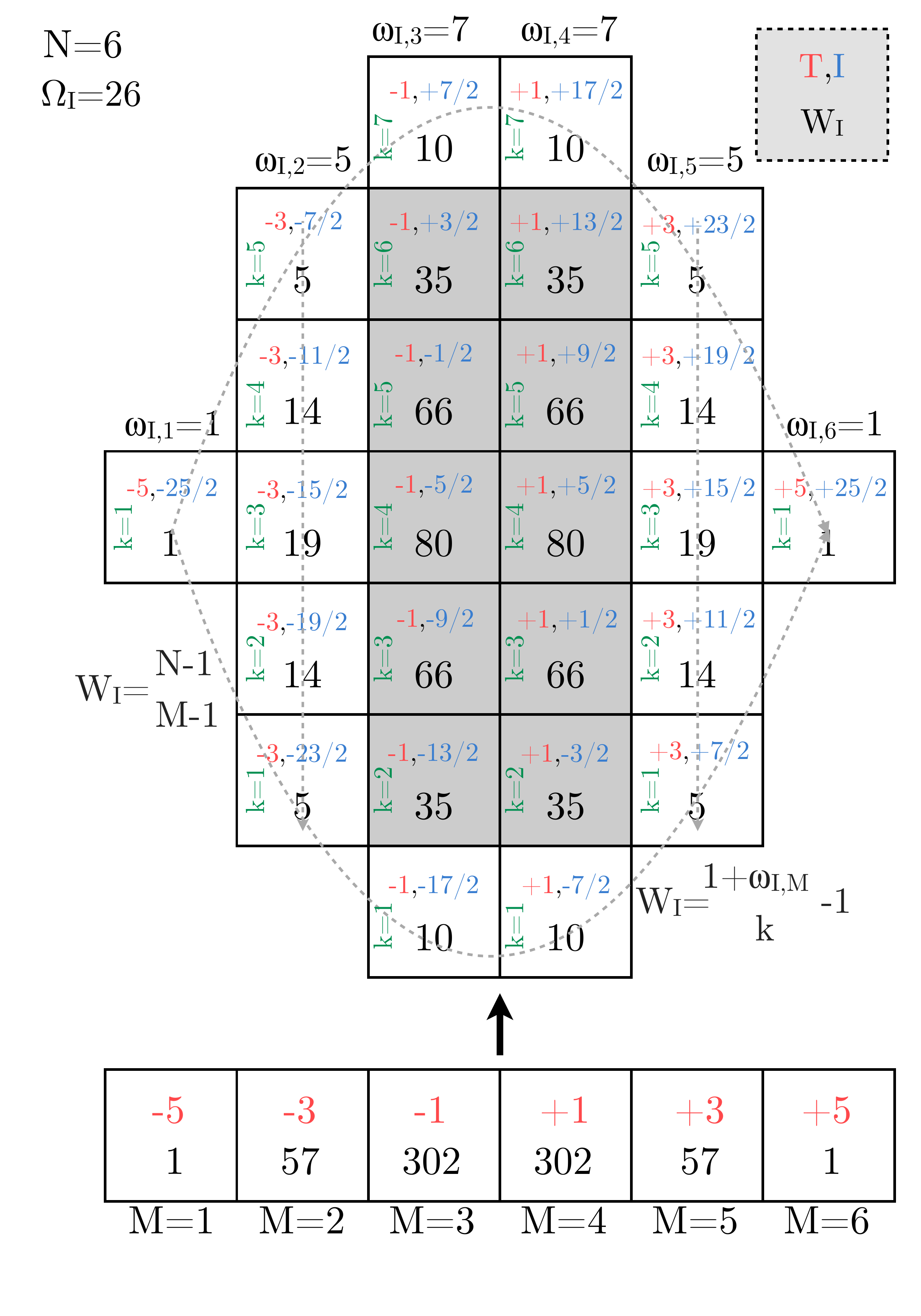}
\caption{
Schematic example of how a list of microstates based on tally-scoring
(bottom row) is split into sub-microstates using the integral-path
based entropy system. The dashed lines depict sub-microstates where
we are able to find analytic formulae to predict the occupancies,
and the shaded sub-microstates are those for which no analytic
solution was found and thus we rely on pre-computed libraries.
}
\label{fig:integral_example}
\end{center}
\end{figure}

On a finer scale, the $M^{\mathrm{th}}$ microstate is expanded into
$\omega_{I,M} = [ (N-M)(M-1) + 1 ]$ sub-microstates. This was
identified by noting that the sequence follows the rascal triangular number
sequence. When summed over $M=1$ to $M=N$, this yields the expected result of
$\Omega_I=\sum_{M=1}^N \omega_{I,M} = \mathcal{C}_{N-1}$ total number of
sub-microstates. In Figure~\ref{fig:integral_example}, we show an example
of how an array of microstates is split into $\Omega_I$ sub-microstates
in the case of $N=6$.

\subsection{Determining Sub-Microstate Index}

Before we deal with the occupancies of each sub-microstate,
we first require a means to convert an observed pair of
scores, $\{T,I\}$, into a sub-microstate index, given by
the numbers $M$ and $k$. The index $M$ is still determined
using the same procedure as before, i.e. it is given soley
by the tally $T$ and Equation~(\ref{eqn:orientT}). To complete
the picture then, we need a means to convert an observed
integral score, $I$, into an index $k$.

To do this, we follow a similar procedure to that used in
Section~\ref{sub:whereami} and first write down the
forward-case of the integral as function of $k$, which
we find can be written as

\begin{equation}
I(k) =
\begin{cases}
2k-I_{\mathrm{max}}		& \text{if } M \leq \tfrac{N}{2},\\
I_{\mathrm{max}}-2k		& \text{if } M > \tfrac{N}{2}.\\
\end{cases},
\label{eqn:Iscores}
\end{equation}

where $I_{\mathrm{max}}$ is the maximum $I$ score for
each index $M$, given by

\begin{align}
I_{\mathrm{max}} &= \tfrac{1}{2} (N-1)^2 - (\mathrm{Min}[M-1,N-M])^2.
\end{align}

In practice, we convert an $I$-score into $k$ by writing out the
full list of possible $I$-scores for each $k$ with Equation~(\ref{eqn:Iscores})
and then selecting the matching example.

\subsection{Occupancies}

The occupancy of these sub-microstates, which we denote using the symbol $W_I$,
does immediately appear to follow a well-known number sequence, but we do
observe that the extreme values of $W_I(M,k=1)$ follows Pascal's triangular
number sequence along the axis $M=1$ to $M=N$. The same is true for the
opposite extreme of $W_I(M,k=\omega_{I,M})$, such that

\begin{align}
W_I(M,k=1||\omega_{I,M}) &= \binom{N-1}{M-1} = \frac{ (N-1)! }{ (M-1)! (N-M)! }.
\end{align}

We also know the sum of the occupancies across all sub-microstates for a fixed
choice of $M$ must equal the $M^{\mathrm{th}}$ microstate's occupancy, such
that

\begin{align}
\sum_{k=1}^{\omega_{M,I}} W_I(M,k) &= A_{N,M-1} \nonumber\\
\qquad&= \sum_{j=0}^{M} (-1)^j \binom{N+1}{j} (M-j)^N,
\end{align} 

where the above illustrates the two sums above cannot be directly equated since
they use different summation indices. 

Thirdly, we note that the second columns, $M=2$ and $M=(N-1)$ appear to follow
a well-known sequence, specifically the Pascal's triangle -1, such that

\begin{align}
W_I(M=2,k) =& \binom{1+\omega_{I,M}}{k} - 1\,\,\forall\,\,k\in\{1,2,...,\omega_{I,M}\},
\end{align}

where we write the formula in two equivalent ways and note that
$W_I(M=2,k)=W_I(M=\omega_{I,M}-1,k)$.

Aside from these specific cases, we are unable to find a general analytic
form for the $W_I(M,k)$ and thus $W_I(T,I)$, Instead, we numerically computed
all possible permutations of 1 to 10 planet systems, counted the occupancies
of each sub-microstate and then saved them to a library function. In instances
where the aforementioned specific cases hold, we employ the analytic solution
instead. This \python\ code is made available at \wwwcoolworlds.

\subsection{Properties}

As noted, including integral path as a second degree of freedom lowers the
entropy of the Solar System. Further, as with the tally-based entropy, we
verified that random swapping leads to ensemble increasing in entropy over
time, as depicted in Figure~\ref{fig:history}. As can also be seen from this
figure and using the equations above, that the total number of unique
microstates is higher with this definition, leading to a finer array of
possible entropies and thus a lower maximum entropy in an absolute sense.

\section{ENTROPY USING CHANGE POINTS}
\label{sec:entropyC}

\subsection{Number of Unique Sub-Microstates}

In addition to integral paths, we devised a alternative way to define the
second degree of freedom based on the number of ``change points'' which occur
in the tally history. This essentially serves like a derivative tally-layer,
where we append $+1$ if $t_{i+1}$ is different from $t_i$, or $+0$ otherwise.
We may write the change point tally, $C$, as

\begin{align}
C &= \sum_{i=1}^{N-2} (1 - \delta[t_i,t_{i+1}])
\end{align}

where $\delta[p,q]$ is the Kronecker Delta function and now
each unique sub-microstate is now uniquely defined by $\{T,C\}$.
To distinguish from before, we label the occupancy of each sub-microstate
with the notation $W_C$.

As before, we find that this definition is effective at splitting up the
originally defined microstates into a finer grid of sub-microstates and again
brings the Solar System down into a lower entropy state, from $W_T=15619$ to
$W_C = 3472$. When replacing Mars with a Super-Earth to create an
``idealized'' Solar System, the difference is greater, going from $W_T=15619$
to $W_C = 70$ (see Figure~\ref{fig:Solar}).

When using change points, the $M^{\mathrm{th}}$ microstate is divided into
$\omega_{C,M}$ sub-microstates, where

\begin{align}
\omega_{C,M} &= 2 \mathrm{Max}\big[\mathrm{Min}[M-1,N-M],\tfrac{1}{2}\big] - \delta_{\tfrac{N-1}{2},M-1}.
\label{eqn:smallomega_m}
\end{align}

Summing over all $N$ and simplifying, we find that the number of unique
microstates grows from $\Omega_T=N$ to

\begin{align}
\Omega_C &= \floor{((N-1)^2+3)/2}.
\end{align}

\begin{figure}
\begin{center}
\includegraphics[width=8.4cm,angle=0,clip=true]{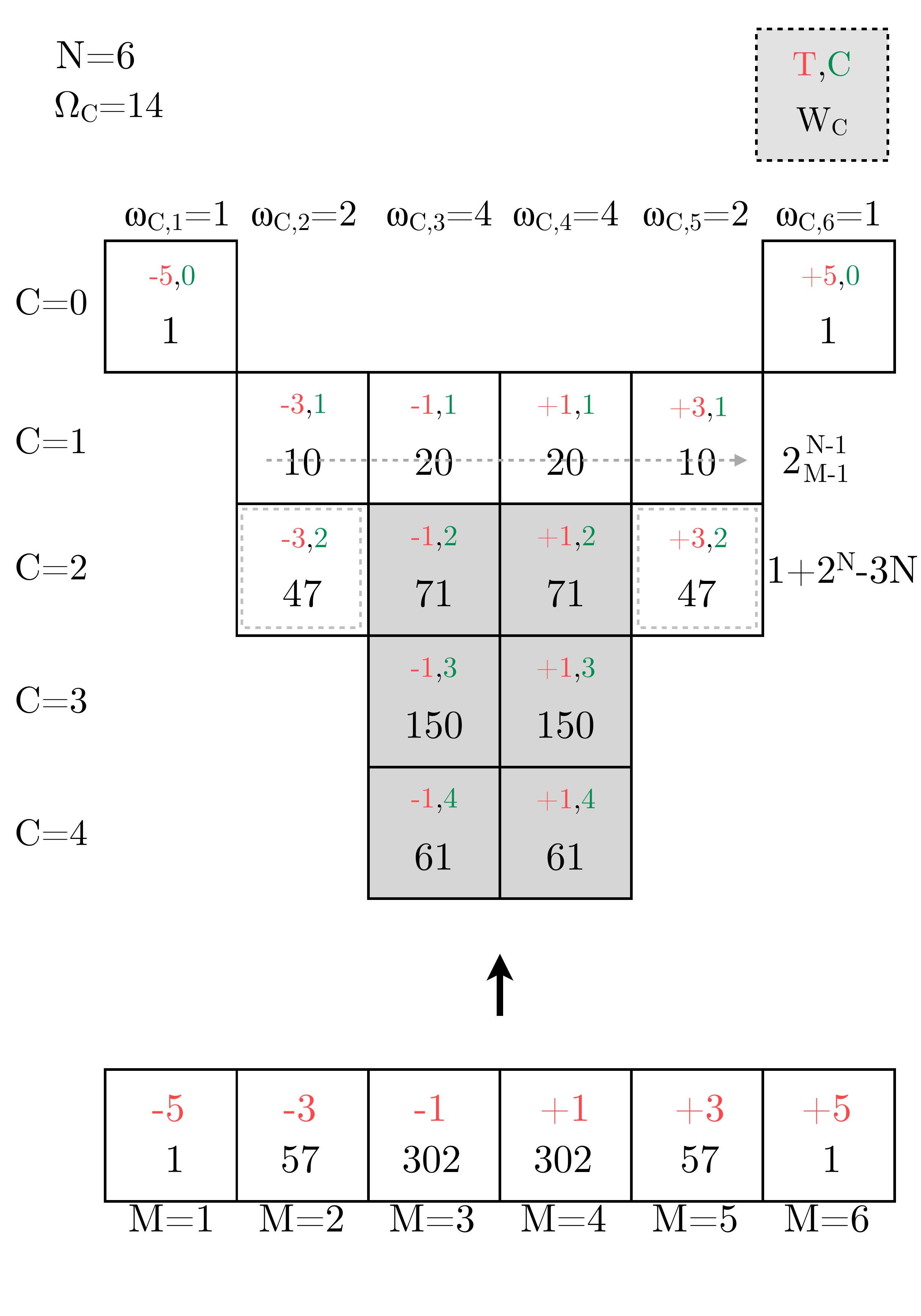}
\caption{
Schematic example of how a list of microstates based on tally-scoring
(bottom row) is split into sub-microstates using the change point
based entropy system. The dashed lines/boxes depict sub-microstates where
we are able to find analytic formulae to predict the occupancies,
and the shaded sub-microstates are those for which no analytic
solution was found and thus we rely on pre-computed libraries.
}
\label{fig:change_example}
\end{center}
\end{figure}

In Figure~\ref{fig:change_example}, we again show an example
of how an array of the tally-based microstates is split into
$\Omega_C$ sub-microstates using the change point system.

\subsection{Determining Sub-Microstate Index}

We briefly point out that unlike the integral-scoring system, the sub-microstate
index for change points is defined by the change point score itself without
manipulation (e.g. see Figure~\ref{fig:change_example}). For this reason, there
is no issue with having to perform a conversion here.
\\
\\

\subsection{Occupancies}

The occupancies of these sub-microstates, $W_C$, does not immediately appear
to follow a well-known number sequence, but we do observe that the extreme
values of $\lim_{C\to0} W_I=1$, where incidentally $C=0$ can only occur
for $M=1$ or $M=N$.

We also find that the $C=1$ row follows a binomial-like pattern,
specifically

\begin{align}
\lim_{C\to1} W_C &= 2 \binom{N-1}{M-1},
\label{eqn:WClim1}
\end{align}

where we note that $C=1$ occurs for all $M$ where $M\neq 1$ or $M \neq N$.

We also know the sum of the occupancies across all sub-microstates for a fixed
choice of $M$ must equal the $M^{\mathrm{th}}$ microstate's occupancy, such
that

\begin{align}
\sum_{C=0}^{\omega_C} W_C(M,N,C) &= A_{N,M-1} \nonumber\\
\qquad&= \sum_{j=0}^{M} (-1)^j \binom{N+1}{j} (M-j)^N.
\end{align} 

Note that, in general, $\omega_{C,M} \neq M$, such that the summation
indices are distinct in the above and thus this does not provide a candidate
general formula for $W_C$.

Finally, for $M=2$ or $M=N-1$, Equation~(\ref{eqn:smallomega_m}) implies that
$\omega_{C,M}(C=2) = \omega_{C,M}(C=N-1) = 2$ for all $N$. Therefore, since we
know $W_C(C=1)$ (Equation~\ref{eqn:WClim1}), and we know the sum of the
occupancies across all sub-microstates for any $M$, then we can write

\begin{align}
W_C(T,C=2) =& \Bigg[\sum_{j=0}^{M} (-1)^j \binom{N+1}{j} (M-j)^N\Bigg] \nonumber\\
\qquad& - \lim_{M\to2||N-1} \Bigg[ 2 \binom{N-1}{M-1} \Bigg],\nonumber\\
\qquad&= 1 + 2^N - 3N.
\end{align}

Beyond this result, we are unable to find any other analytic formula to
expedite the calculation of occupancies. Instead, and as before, we ran
numerical experiments exploring all permutations of planetary size orderings
for each $N$ up to $N=10$ and then simply counted the occupancies. These
results were coded up into a \python\ library (see \wwwcoolworlds), such that
for any given $T$ and $C$ combination, one can simply look-up the corresponding
$W_C$, leveraging the analytic results from above wherever possible.
\\
\\

\subsection{Properties}

As with the integral path, including change points as a second degree of
freedom decreases the entropy of the Solar System. Further, as with both
previous systems, we again find that random swapping leads to ensemble
increasing in entropy over time, as expected and depicted in 
Figure~\ref{fig:history}. As with the other definitions then, the observation
of a low entropy state can be stated to be unlikely to arise from random
swaps over time and the observation of an ensemble of systems with low
entropy is highly unlikely to arise from random swaps. In other words,
such low entropy states are fundamentally not random but contain some
information about a guiding process leading preferentially to such low
entropy configurations.

\section{APPLYING TO KEPLER SYSTEMS}
\label{sec:kepler}

\subsection{Interpreting Entropy}

In this section, we apply our entropy scoring system to real planetary systems.
Before doing so, we briefly highlight some conclusions which can be made using
an entropy score.

For an individual system found to have a low entropy (using any of previously
discussed metrics), the $p$-value of observing such an entropy under the
hypothesis that all systems are randomly organized can be computed. By 
``randomly organized'', we specifically refer to the simulations introduced in 
Section~\ref{sub:evolve} i.e. initially zero-entropy systems allowed to undergo
a large number of position exchanges. The $p$-value computation is performed by
simply evaluating the median\footnote{to account for duplicates} rank of the 
observed entropy within the sorted list of Monte Carlo simulated final states
for random systems.

To give an example of the above, consider the $N=5$ planet system Kepler-80
\citep{kepler80}, which was depicted earlier in Figure~\ref{fig:example}.
Kepler-80 exists in a perfectly ordered configuration, giving $S_T=S_I=S_C=0$.
Even from a sample of $10^6$ randomly generated systems, we find 0 instances of
such a low entropy for any of metric and thus infer that $p_T, p_I, p_C \leq 
10^{-6}$.

The $p$-values quoted, a measure of surprisingness, reveal that even amongst
a sample of several thousand planetary systems, such a low entropy score is
not expected. We attribute this as evidence that Kepler-80 contains a memory
of a specific formation pathway which presumably is described by some
unknown entropy distribution with a greater probability density at low
entropy values. In plainer terms, Kepler-80 appears to remember something of
its origin.

Ideally, this process could be repeated on all of the \textit{Kepler} systems.
However, the argument above has ignored measurement uncertainties and this
introduces a major obstacle to realizing an entropy for each system. In many
cases, the measurement uncertainties reported on the NASA Exoplanet Archive
(\href{http://exoplanetarchive.ipac.caltech.edu/}{NEA}; \citealt{NEA}) are
sufficiently wide that one should expected a significant fraction of the joint
posterior samples to lead to distinct entropy scores. Essentially, this implies
we need to derive a posterior for the entropy.

Marginal posterior distributions of each planet's ratio-of-radii could be used
to achieve this. However, we are aware of no such public catalog at this time
that treats planets with the same parent star as transiting a star with a global
set host star parameters and covariant individual planet parameters
(\citealt{sandford2017} present such posteriors but only for a subset of
\textit{Kepler} stars). Instead, we generate representative posteriors using the
method described in what follows.

\subsection{Generating Asymmetric Posteriors}
\label{sub:postgen}

We first downloaded the list of reported ratio-of-radiis from \href{http://exoplanetarchive.ipac.caltech.edu/}{NEA} \citep{NEA} for every
\textit{Kepler} system with $N\geq3$ planets, for which the host star satisfied
$\log g>4$ and has an exoplanet disposition of being either candidate or
confirmed (224 systems). Typically, the ratio-of-radii are reported with
asymmetric measurement uncertainties, meaning that one cannot simply treat them
as being described by a normal distribution, which is symmetric.

\begin{figure}
\begin{center}
\includegraphics[width=8.4cm,angle=0,clip=true]{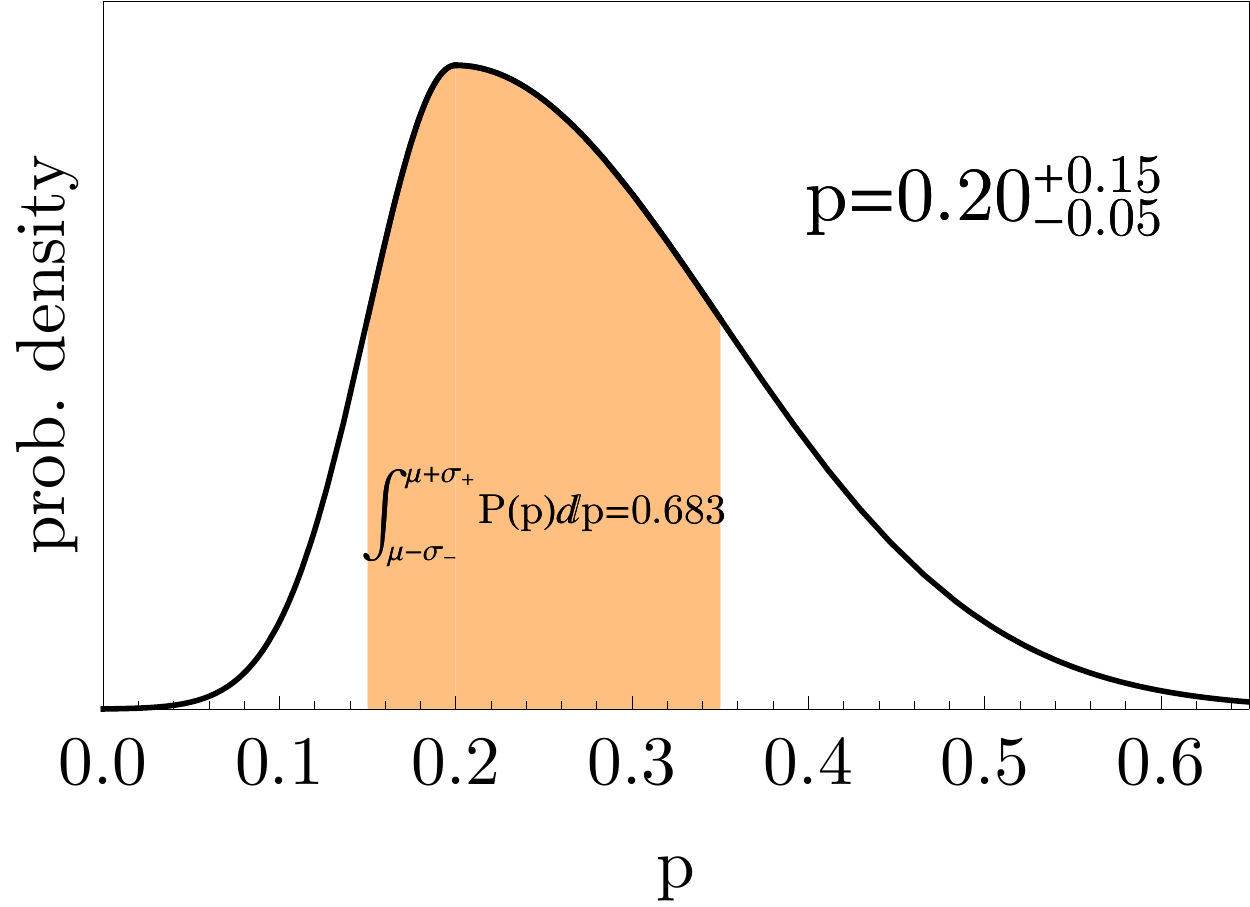}
\caption{
Example of an asymmetric probability distribution generated using our
weighted mixture model of truncated normals.
}
\label{fig:pdfexample}
\end{center}
\end{figure}

To create an asymmetric distribution, we used a mixture of two truncated normals,
where the first normal is truncated from $0$ up to the reported mean, and the
second is truncated from the reported mean up to unity. To ensure a smooth
probability distribution at the boundary, the mixture weights are set to the
ratio of each density at the mean's location, leading to a normalized and smooth
asymmetric distribution. Specifically, for a measurement of
$p=\mu_{-\sigma_{-}}^{+\sigma_{+}}$, our model treats $p$ as being distributed as

\begin{equation}
p \sim
\begin{cases}
\mathcal{T}[\{0,\mu\},\mathcal{N}[\mu,\sigma_{-}]]  & \text{if } p \leq \mu ,\\
\mathcal{T}[\{\mu,1\},\mathcal{N}[\mu,\sigma_{+}]]  & \text{if } p > \mu ,\\
0  & \text{otherwise},
\end{cases}
\end{equation}

where $\mathcal{N}[a,b]$ is a normal distribution and
$\mathcal{T}[\{\alpha,\beta\},\mathcal{X}]$ is a truncated distribution of
$\mathcal{X}$. The resulting probability density function is plotted in
Figure~\ref{fig:pdfexample} for some example inputs.

We performed inverse transform sampling of the distribution to generate
$10^4$ fair realizations of each planet's ratio-of-radii. We then repeat
our calculation of the size-ordering entropy on each realization to
build an entropy posterior for each system. In the case of Kepler-80,
we find that the entropies are measured to be $S_T = 0_{-0}^{+4.2}$\,nats,
$S_I = 0_{-0}^{+2.8}$\,nats and $S_C = 0_{-0}^{+3.1}$\,nats, which
illustrates the considerable effect of current measurement uncertainties.

Critically, our model does not account for covariance which would lead to
tighter constraints on the resulting entropy posteriors. Ultimately, our model
is therefore accurate but less precise than possible and future work could
revisit the calculations described below when covariant posteriors become
available. 

\subsection{Application to Kepler Multis}
\label{sub:kepapp}

\begin{figure*}
\begin{center}
\includegraphics[width=18.4cm,angle=0,clip=true]{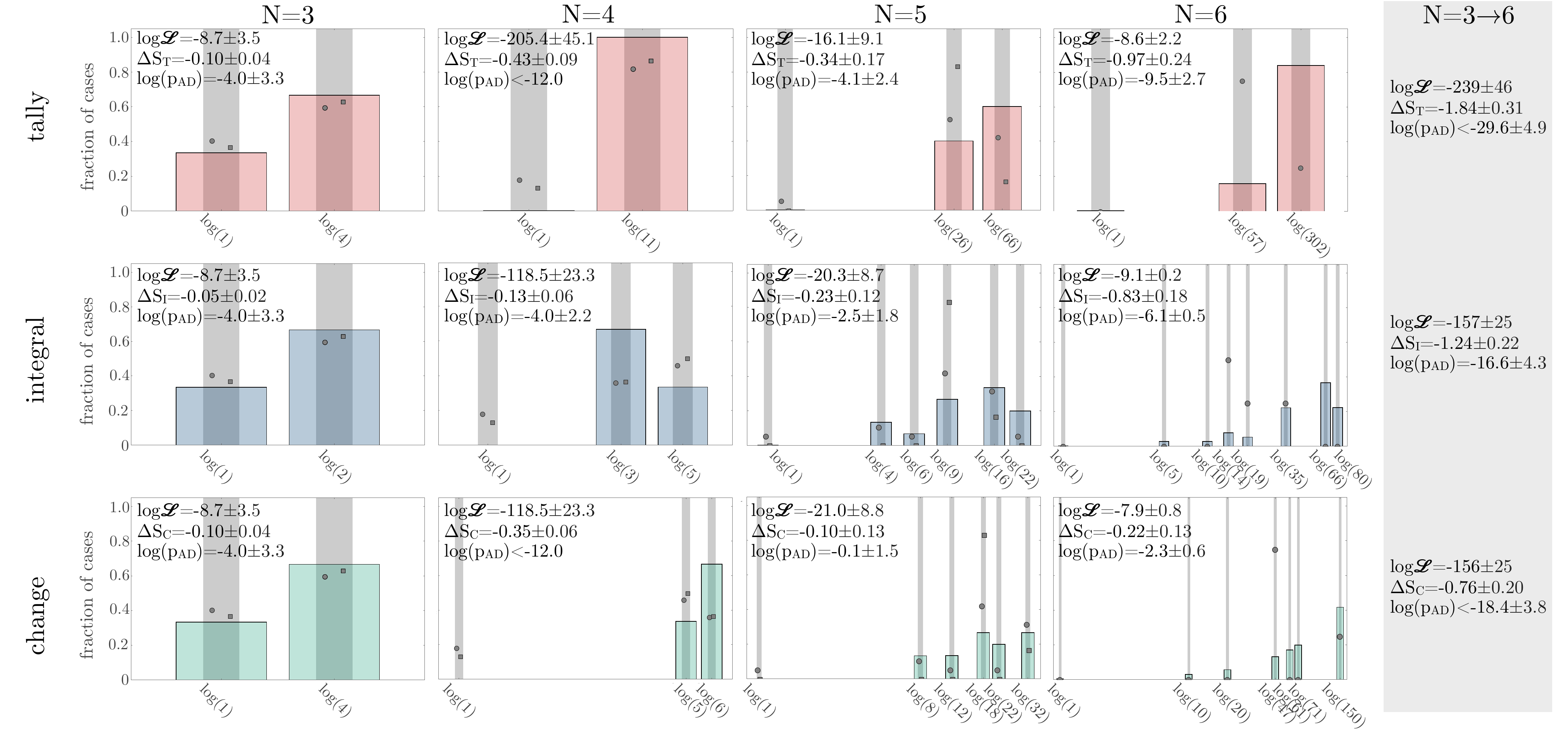}
\caption{
Summary of results comparing the entropy of multiple planet \textit{Kepler}
systems (columns) for three different entropy-scoring methods (rows).
Final column adds up the multis together, reflecting the consistent
pattern of \textit{Kepler} multis (circles with Poisson counting error
bars) having a lower entropy than random systems (bar histograms).
Squares points are the same \textit{Kepler} systems but
filtering on only those thought to be dynamically packed.
}
\label{fig:kepler}
\end{center}
\end{figure*}

We applied our algorithm to the 224 \textit{Kepler} multi-planet systems
described earlier. Grouping the systems into their unique $N$ values, we find
151 three-planet systems, 50 four-planet systems, 19 five-planet systems and
4 six-planet systems. For each system, we computed all three entropy scores
and then created histograms of the resulting distributions of the
best-reported ratio-of-radii, shown in Figure~\ref{fig:kepler}.

Each panel in Figure~\ref{fig:kepler} is for $N=3,4,5,6$ across and
$S_T,S_I,S_C$ down. In the top-left of each panel, we report three
summary statistics of interest, where the best-reported value is that
derived using the best-reported ratio-of-radii, whereas the ``$\pm$''
uncertainty is the standard deviation of each metric across $10^4$
random posterior draws of the ratio-of-radii, as described earlier.

The three metrics considered are $\log\mathcal{L}$ assuming a binomial
distribution at each unique microstate, the mean entropy of the
population minus that of a random population, $\Delta S$, and the $p$-value
derived from comparing the observed population with a random population
using the Anderson-Darling test, $\log(p_{\mathrm{AD}})$.

Across the board, we consistently find that the \textit{Kepler} population
has a lower entropy than the random population for all $N$ and entropy
scoring systems, although the magnitude and significance varies considerably.
Combining the metrics for all planet groupings on the far-right column of
Figure~\ref{fig:kepler} shows that there is very strong evidence for an
entropy deficit amongst the \textit{Kepler} systems, ranging from at best
22\,$\sigma$, using the tally-based binomial likelihood, to at worst
$3.8$\,$\sigma$, using the mean entropy of the change-point entropy. The
Anderson-Darling tests typically sit between two extremes, suggesting a
$\sim 6$\,$\sigma$ effect.

Considering each $N$ grouping separately, we note that $N=4$ group seems to
show a particularly pronounced entropy deficit, so extreme that we were unable
to compute numerically stable Anderson-Darling $p$-values. In general, the
trend appears to be that $N=3$ systems show a modestly significant deficit, it
becomes extreme for $N=4$ and then drops down with increasing $N$. This downward
trend is likely due to the ever-smaller samples sizes with increasing $N$ which
naturally attenuate significances. The $N=3$ case may be explained by the fact
that only two unique microstates are possible for all three entropy-scoring
systems and random systems tend to populate both with non-negligible fractions.

Since the integral path system does the best job of explaining the Solar
System, we tend to prefer it in this work. Accordingly, we conclude that there
is $(5.4\pm0.8)$\,$\sigma$ evidence for an entropy deficit in the ensemble of
\textit{Kepler} multis using the Anderson-Darling test, and a consistent
$5.6$\,$\sigma$ confidence when using the means testing. This analysis
indicates that the ensemble of \textit{Kepler} multi-planet systems exist in a
lower entropy state than that of pure randomization. In other words, the
\textit{Kepler} systems contain some information or memory of their
(non-random) origin mechanism, since they are highly unlikely to have arrived
in their observed configuration as a result of random exchanges. In many ways
the above statement may sound completely expected, yet we have shown here that
the statement is non-trivial to formally prove.

\subsection{Dynamically Packed Systems}

Our definition of entropy is sensitive to missing planets. An obvious set of
missing planets are those beyond 1\,AU, where transit surveys like
\textit{Kepler} have weak sensitivity \citep{2008ApJ...686.1302B,
2016MNRAS.463.1323K}. However, the entropies measured interior to 1\,AU are
still valid when treated as the local entropy score of this region. A much
more problematic situation is the case of one or more missing planets interior
to the outer-most detected planet but exterior to inner-most detected planet.
These worlds will cause even this local entropy score to be erroneous and thus
we consider how to mitigate against this effect here.

However, we first highlight that the \textit{Kepler} multi-planet systems have
been demonstrated to exhibit low mutual inclinations (consistent with
$\lesssim3^{\circ}$; \citealt{2011Natur.470...53L,2012AJ....143...94T,
2012ApJ...761...92F,2014ApJ...790..146F,2016ApJ...816...66B}), which makes it
geometrically improbable for a planet interior to the outermost transiting planet
to be missing i.e. non-transiting. Much more likely is that the planet is
missing due to detection effects, most plausibly because the planet is so small
it evaded detection.

To investigate this, we may use the generalized Titius-Bode law
\citep{bovaird2013} as a proxy for dynamical packing, for the following
reasons. Consider a $N$-planet system which is dynamically packed, such that
the planets are so close together that additional planets cannot be injected
into intermediate orbits. Such a configuration implies that the planets are
separated by just $k$ Hill radii, where $k$ is some number greater than one
controlling the spacing and the Hill radius of the $i^{\mathrm{th}}$ planet,
$H_i$, is given by

\begin{align}
H_i &= a_i \Big( \frac{\mathcal{M}_i}{3} \Big)^{1/3},
\end{align}

where $\mathcal{M}_i$ represents the mass ratio between the $i^{\mathrm{th}}$
planet and the star. 

For $N=2$ planet systems, $k=2$-$4$ is expected to ensure stability
\citep{1984Icar...60...40W,1985Icar...63..290W,1987Icar...69..249L,
1988merc.book..670W,1993Icar..106..247G} but for $N>2$, no choice of $k$ is
indefinately stable, rather the stability time increases with $k$ such that
$k\gtrsim13$ provides Gyr stability to a Solar System analog
\citep{1996Icar..119..261C}. For our purposes, it is unimportant what value $k$
actually is, but let's proceed to treat it is a fixed number. Accordingly,
semi-major axes of the $i^{\mathrm{th}}$ and $(i+1)^{\mathrm{th}}$ planet,
$a_i$ and $a_{i+1}$, must satisfy

\begin{align}
a_{i+1} + k H_{i+1} &\geq a_{i} - k H_i,
\end{align}

which in the limit of the equality gives

\begin{align}
a_{i+1} &= a_i \frac{ 1 + 3^{-1/3} k \mathcal{M}_i^{1/3} }{ 1 - 3^{-1/3} k \mathcal{M}_{i+1}^{1/3} }.
\end{align}

Converting from semi-major axes to periods via Kepler's Third Law
(and assuming $\mathcal{M}_j\ll1$) gives

\begin{align}
P_{i+1} &= P_i \Bigg( \frac{ 1 + 3^{-1/3} k \mathcal{M}_i^{1/3} }{ 1 - 3^{-1/3} k \mathcal{M}_{i+1}^{1/3} } \Bigg)^{3/2},
\end{align}

or

\begin{align}
\log(P_{i+1}) &= \log(P_i) + \underbrace{\tfrac{3}{2} \log \Bigg( \frac{ 1 + 3^{-1/3} k \mathcal{M}_i^{1/3} }{ 1 - 3^{-1/3} k \mathcal{M}_{i+1}^{1/3} } \Bigg)}_{=\log\alpha}.
\label{eqn:logalpha}
\end{align}

This result essentially states that in the limit of low mass ratios,
planets separated by $k$ Hill radii will follow the generalized
Titius-Bode (GTB) law of \citet{bovaird2013}, i.e. that

\begin{align}
\log P_i &= P_1 + (i-1) \log \alpha.
\end{align}

This result is consistent with that found by \citet{hayes1999}, who
use numerical integrations to demonstrate that the GTB law is simply a
consequence of dynamical stability of packed systems.

We may now use the GTB law as a simple proxy to detect dynamically packed
systems, without conducting detailed N-body tests such as those presented in
\citet{fang2013}. To accomplish this, we compare how well $\log \alpha$ agrees
between consecutive pairs of planets in a system. If $\log \alpha$ is
consistent amongst all pairs, then it can be said that the GTB law holds and
the planets are consistent with being dynamically packed.

This criterion requires some quantification since small deviations in
$\log\alpha$ are to be expected by varying the mass ratios, $\mathcal{M}_i$,
and $k$, as can be seen from Equation~(\ref{eqn:logalpha}). Accordingly,
we evaluated the range of $\log\alpha$ values expected as a function of $k$
for mass-ratio ranges from 0.1 - 10\,$M_{\oplus}$, consistent with the
typical \textit{Kepler} planets. We find that the fractional variation
in $\log\alpha$ rises from $\simeq \tfrac{1}{4}$ for $k=1$ to
$\simeq \tfrac{1}{3}$ for $k=20$. We therefore elect to use a third as a
tolerance level to test for.

Applying this criterion reduces the samples sizes slightly. Specifically,
we find go from 151 to 103 three-planet systems, 50 to 30 four-planet systems,
19 to 6 five-planet systems and 4 to 0 six-planet systems.

After filtering, we repeated the analysis described in 
Section~\ref{sub:kepapp} for each $N$, and the points are plotted in
Figure~\ref{fig:kepler} as squares (except for $N=6$ where no packed
systems were identified). The results and metrics are broadly consistent
with slightly deflated significances due to the smaller sample size
under investigation. We find that the total entropy differences between
the \textit{Kepler} $N=3\to5$ planet systems and a randomly generated
population are $\Delta S_T = -0.77\pm0.38$, $\Delta S_I = -0.22\pm0.27$
and $\Delta S_C = -0.26\pm0.30$. The Anderson-Darling test again supports
strong evidence for a distinct population, with $p$-values of 5.8, 5.1 and
4.7\,$\sigma$ for the tally-, change- and integral-based entropies.

\subsection{Comparison to other approaches}

%
The entropy methods devised and applied in this work are tailored to our
specific problem in mind, yet we highlight that there is a large prior
literature for checking randomness. A classic example is the Wald-Wolfowitz
runs test, which is a non-parametric tool for evaluating the randomness of a
binary sequence, allowing one to test the hypothesis that the elements of a
sequence are mutually independent \citep{bradley:1968}. The sequence of
planetary radii within a system is, of course, not a binary sequence but rather
represented by a set of continuous, real numbers. However, we may apply the
runs test to the tally-scores, which asks whether the next member is larger or
smaller than the previous.

The runs test, therefore, can be a useful tool in the context of the
evaluating the surprisingness of the tally-based entropy method's
resulting tally sequence\footnote{
We also highlight that the standard implementation of the runs test invokes the
central limit theorem to approximate the distribution for number of runs as
being normal, but in our case the number of planets around each star is small
and so this condition is violated.
}.
However, it does not delineate the results into distinct classes nor quantify
the information content, in the same way an entropy scheme can (although we
highlight that contemporary work has tried to connect the classic runs test to
entropy scores; e.g. \citealt{rukhin,kraskov:2004,gan:2015}). Going further,
the test was obviously not designed with the planet problem in mind, and does
not capture our physical insights that the lowest-entropy systems should be
described by just two sequences, as our integral-based aims to account for.

%
Another common approach we highlight is the binary entropy function, $H_b$,
which returns the entropy of a Bernoulli process of probability, $p$:

\begin{align}
H_b &= -p \log p - (1-p) \log (1-p),
\end{align}

where using natural log returns an entropy score in nats. For example, a
fair coin with $p=0.5$ has the maximum possible entropy score of
$H_b=1.44$\,nats. At a basic level, this entropy score is
fundamentally different from those considered in this work since we do not
assume that the tally scores derived from a sequence of planetary radii follow
a Bernoulli distribution. Rather, we simply compute the full range of
possible configurations, assign entropies based on proxy scores and
associated microstate occupancies, and then test for significance using Monte
Carlo experiment. In the binomial entropy context, the entropy is used as a
measure of uncertainty about a process or event, whereas our entropy scores
are designed to be a measure of randomness.

%
We also highlight that there exists a suite of contemporary tests for
quantifying the entropy of random number generators used in computer
simulations, such as the Diehard tests\footnote{See
http://webhome.phy.duke.edu/~rgb/General/dieharder.php}, but these generally
focus on very long sequences (unlike considered here) and, as before, were
not designed to capture physical intuition for planetary systems.

\section{DISCUSSION}

We have presented three different formalisms for evaluating the entropy of
planetary systems, in terms of their size orderings. Inspired by the stark
contrasts between ostensibly ordered systems, like Kepler-80 \citep{kepler80},
and disordered systems, like Kepler-20 \citep{kepler80}, our work aims to 
provide a quantitative framework for evaluating these evident qualitative
differences.

Using a tally-based scoring system, an integral-path method and a change-point
system, we show that all three have marginal entropies that increase with
respect to a proxy for time, as should be expected. We provide a detailed
mathematical account of our definitions and show that much, but not all, of
the microstate occupancies can be expressed with closed-form solutions,
enabling fast computation. Cases without such solutions are folded in using
look-up tables, culminating in our public \python\ package for evaluating the
different entropies at \wwwcoolworlds.

Through Monte Carlo simulation, we predict the expected distribution of the
entropies for various $N$-planet systems after they have evolved from a large
number of random swaps. When comparing this randomly-generated distribution to
that of the real \textit{Kepler} systems, we consistently find an entropy
deficit in the real data to high confidence. Since the \textit{Kepler} systems
exhibit lower entropy than that expected of pure random swaps, the origin
of their entropy values is highly unlikely to be random. In other words, they
must contain some information or memory about the specific conditions which led
to their original configuration, which may have been partially eroded by
subsequent dynamical evolution. Nevertheless, the formal demonstration that the
size-ordering of \textit{Kepler} multis contains information establishes that
efforts to infer initial formation conditions are not necessarily in vein.

We highlight that there is much room for improvement upon our proposed entropy
schemes. First, planets of nearly equal sizes versus planets of vastly different
sizes will yield the same entropy score for the same size ordering. We encourage
future work to investigate the value of adding in a Boltzmann-like constant
in front of our entropy definition, for which an obvious candidate is the
variance of the sizes. Second, our definition does not address the specific
distribution of orbital separation, merely their rank order. It may be possible
to add such a term, although this is complicated by the dynamical preference for 
near mean-motion resonances of various orders.

Finally, we have not investigated the possibility of entropy differences
between different sub-populations. Since entropy is expected to increase with
respect to time, it should behave as an age-proxy, although it is unclear
whether the time-scale for significant entropy evolution is comparable to the
typical ages of observed systems. Regardless, it would be worthwhile to
investigate if young versus old stars harbor different entropies. Similarly,
one might expect binary stars to induce greater dynamical mixing and thus
lead to high entropies. Such investigations are non-trivial since a complete
and precise catalog of stellar ages and binarity do not presently exist for
the \textit{Kepler} sample, nor do mulit-planet covariant transit posteriors
necessary for precise population comparison work. For these reasons, we
highlight these as possible objectives for future work at this time.

As a concluding remark, we hope that this work encourages research into the
relativity young discipline of ``exo-informatics'', a field which can
complement the broader population-based statistical inference approaches coming
into play.

\acknowledgments

The author is grateful to Eric Feigelson for helpful comments in revising
this manuscript.
DMK thanks members of the Cool Worlds Lab for lively conversations on the
topic of exoinformatics. Special thanks to Chris Lam and Ruth Angus for helpful
suggestions.
This research has made use of the NASA Exoplanet Archive, which is operated by
the California Institute of Technology, under contract with the National
Aeronautics and Space Administration under the Exoplanet Exploration Program.

\bibliography{mainbib}



\listofchanges

\end{document}